# Fabrication of tetragonal FeSe – FeS alloy films with high sulfur contents by alternate deposition

Kohei Fujiwara[*], Junichi Shiogai, and Atsushi Tsukazaki

*Institute for Materials Research, Tohoku University, Sendai 980-8577, Japan*

E-mail: kfujiwara@imr.tohoku.ac.jp

We report the synthesis of tetragonal FeS$_x$Se$_{1-x}$ films ($x \leq 0.78$) by pulsed-laser deposition. To fabricate the tetragonal alloy films with tetragonal FeSe and hexagonal FeS targets, we adopted an alternate deposition technique with FeSe buffer layer on MgO(001). The overall film composition is controlled by the thickness ratio of FeS / FeSe layers. The out-of-plane lattice parameter of the films follows Vegard's law, demonstrating homogeneous alloying by inter-diffusion. The sulfur solid solubility reaches $x = 0.78$ in the FeS$_x$Se$_{1-x}$ films, which is by far larger than $x \sim 0.40$ in bulk governed by the tetragonal phase instability.

The PbO-type tetragonal FeSe is the simplest compound among Fe-based superconductors, showing superconductivity at a critical temperature ($T_c$) of 8 K.[1] Since its tetrahedrally coordinated Fe plane is the common structural unit of Fe-based superconductors, the electronic structure and paring mechanism of FeSe[2,3] have been one of the central topics in superconductivity research. In addition, the formation of electronic nematic phase[4] and, more recently, the emergence of high-$T_c$ superconductivity in the 1 u.c.(unit cell)-thick[5,6] and ultrathin-film conditions[7,8] have raised further interest in FeSe. The simple crystal structure is also suitable for investigating the chemical substitution effect on the structural and electronic properties. In particular, the alloy system with related chalcogenide compounds[9] have been studied intensively.[10–14] Fabrication of tetragonal alloy films with isostructural FeTe has been reported by pulsed-laser deposition (PLD);[15] $T_c$ was found to rise to 23 K for FeSe$_{0.8}$Te$_{0.2}$ on CaF$_2$ substrate where the phase separation inherent in bulk samples is effectively suppressed by the PLD process.[16] Another chalcogenide alloy candidate is FeS$_x$Se$_{1-x}$.[12–14] Although PbO-type FeS can be obtained by the hydrothermal method ($T_c \sim 5$ K),[17] sulfur solid solubility as tetragonal FeS$_x$Se$_{1-x}$ bulk alloy is as low as $x \sim 0.40$.[10,18] This is primally due to the existence of thermodynamically stable hexagonal FeS phase, which gives rise to a structural deformation at higher $x$.[18] Using PLD in this study, we demonstrate that tetragonal FeS$_x$Se$_{1-x}$ can be stabilized up to $x = 0.78$.

Films were deposited on MgO(001) substrates at 300 °C in a vacuum (a base pressure of the order of 10$^{-6}$ Torr), with tetragonal FeSe and hexagonal FeS polycrystalline targets supplied from Kojundo Chemical Laboratory Co., Ltd. Our preliminary attempts to grow tetragonal FeS directly on MgO(001) at various temperatures, and on other substrates, LaAlO$_3$(001), SrTiO$_3$(001), (La,Sr)(Al,Ta)O$_3$(001), and TiO$_2$(001), were not successful. We therefore employed 2 nm-thick FeSe buffer and subsequent FeS / FeSe alternate deposition to stabilize the tetragonal phase with high sulfur content, as schematically shown in Fig. 1(a). Each layer thickness was tuned by laser pulses irradiated to FeSe and FeS targets. As listed in Table I, the thickness ratio of FeS over the whole film (denoted as FeS thickness ratio hereafter) was varied from 0.2 to 0.8 while each thickness was kept less than 2.2 nm (~4 u.c.) to suppress the hexagonal phase formation of FeS. The total film thickness was approximately 15 nm.

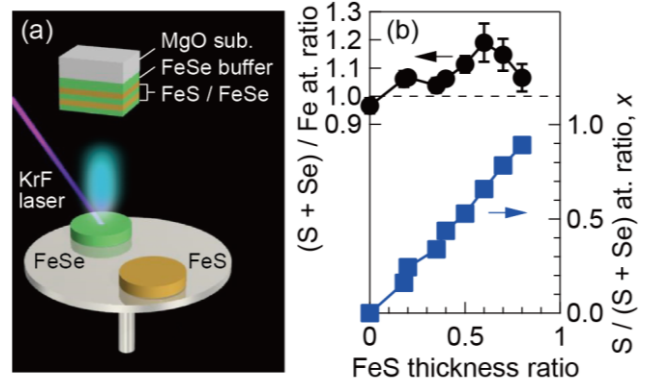

**Fig. 1.** (a) Schematic of alternate deposition of FeS / FeSe stacked films by pulsed-laser deposition. The films were deposited on 2 nm-thick FeSe buffer / MgO(001) substrates. (b) Chemical composition analysis results for (S + Se) / Fe and S / (S + Se) as a function of FeS thickness ratio. Error bars for S / (S + Se) are smaller than the symbols.

Table I. Thicknesses of FeS and FeSe used for alternate deposition. The cycle number of alternate deposition was set to 6. The total film thicknesses including 2 nm-thick FeSe buffer were approximately 15 nm. See text for the definition of FeS thickness ratio and sulfur content $x$.

| Thickness of FeS layer (nm) | Thickness of FeSe layer (nm) | Number of cycles | FeS thickness ratio | Sulfur content, $x$ |
|---|---|---|---|---|
| 0.50 | 1.70 | 6 | 0.20 | 0.24 |
| 1.01 | 1.19 | 6 | 0.40 | 0.44 |
| 1.26 | 0.94 | 6 | 0.50 | 0.53 |
| 1.52 | 0.68 | 6 | 0.60 | 0.66 |
| 1.77 | 0.43 | 6 | 0.70 | 0.78 |
| 2.03 | 0.17 | 6 | 0.80 | 0.89 |

We first evaluated sulfur content in the as-grown FeS / FeSe stacked films by electron probe microanalysis. Figure 1(b) summarizes the chemical composition ratio of (S + Se) / Fe and S / (S + Se) as a function of FeS thickness ratio. The stacked films are slightly anion rich, with approximately 10 ± 5 at% excess anion against Fe cation. This might be due to composition deviation in the ablation process[19)] of FeS because the ratio for FeSe is close to unity. In this work, we focus on the S substitution for Se; we simply express the film composition as FeS$_x$Se$_{1-x}$ and define $x$ as S atomic fraction in the anions, i.e., S / (S + Se). Table I and Fig. 1(b) demonstrate that $x$ in the films can be controlled by the FeS thickness ratio.

Figure 2(a) shows out-of-plane x-ray diffraction (XRD) patterns for FeSe and FeS / FeSe stacked films. FeSe ($x = 0$) single layer exhibits an intense (001) peak at 16.1° associated with clear thickness fringes. In FeS / FeSe stacked films, peaks appear between 16.1° for FeSe(001) and 17.6° for tetragonal FeS(001), which shifts to higher angles with increasing $x$ and disappears at $x = 0.89$. No peaks assignable to secondary phases as well as peak splitting indicative of phase separation are discerned. The absence of crystalline peaks for $x = 0.89$ implies that it is hard to grow tetragonal or hexagonal FeS single layer on FeSe buffer layer. The lattice parameter calculated from those peaks is in good agreement with $c$-axis lengths of bulk tetragonal FeS$_x$Se$_{1-x}$ ($x \leq 0.4$) (Ref. 10) as well as Vegard's law in between bulk values of tetragonal FeSe and FeS (Ref. 17), as plotted in Fig. 2(b). It is therefore reasonable that $c$-axis oriented FeS$_x$Se$_{1-x}$ isostructural to tetragonal FeSe is formed in our films. The linearly decreasing lattice constant holds up to $x = 0.78$ in the FeS$_x$Se$_{1-x}$ alloy films. This is in stark contrast to the bulk result, which clearly deviates from the relation above $x = 0.40$.[10,18)] These observations suggest that the FeS$_x$Se$_{1-x}$ alloy films are obtained from alternately deposited FeSe and FeS stacked films owing to an inter-diffusion, which likely occurs during deposition at 300 °C. Measurements of in-plane lattice parameters and rotational symmetry with much thicker films in future work will provide further support.

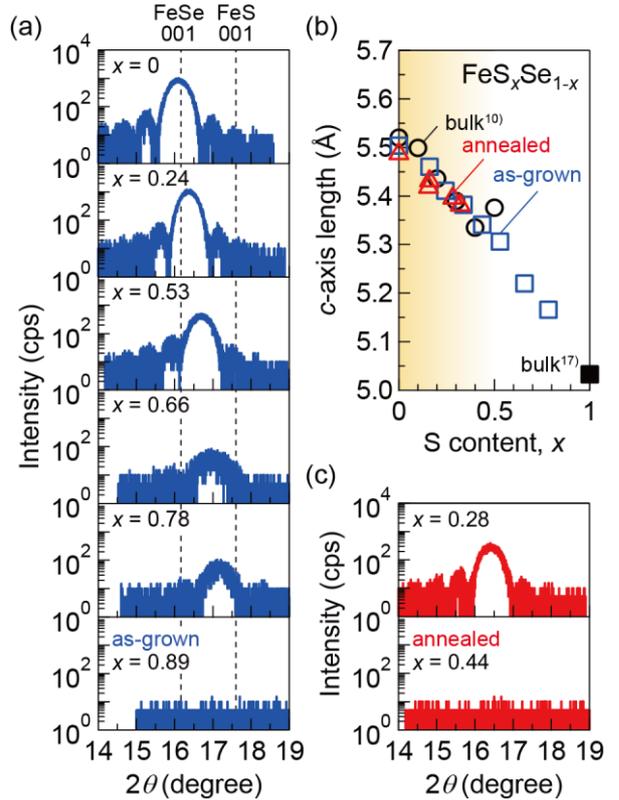

**Fig. 2.** (a) XRD patterns for as-grown FeS$_x$Se$_{1-x}$ films around the tetragonal FeSe(001) and FeS(001) diffraction angles, measured with Cu $K\alpha$1 radiation. (b) S content ($x$) dependence of $c$-axis length in FeS$_x$Se$_{1-x}$ films, the values of which are estimated by assuming the $c$-axis oriented tetragonal phase. Data for bulk tetragonal FeS$_x$Se$_{1-x}$ (Ref. 10) and FeS (Ref. 17) in literature are included. (c) XRD patterns for annealed FeS$_x$Se$_{1-x}$ films.

By *in-situ* annealing FeS$_x$Se$_{1-x}$ films for 30 minutes at 450 °C, we examined the stability of tetragonal alloy films for $x$ below and above the bulk solubility limit. Figure 2(c) shows XRD patterns for annealed FeS$_x$Se$_{1-x}$ films with $x$ = 0.28 and 0.44. Even after annealing, the diffraction peak position for $x$ = 0.28 remains unchanged (also see Fig. 2(b)), evidencing the thermodynamically stable tetragonal phase at low $x$. No additional annealing effect also corroborates the completion of inter-diffusion during deposition. In contrast, the disappearance of (001) peak for $x$ = 0.44 indicates that the high-$x$ alloy decomposes, presumably due to the thermodynamically unfavorable tetragonal phase.[18] Thus, non-equilibrium process of PLD is beneficial to stabilize the tetragonal phase with high sulfur contents.

Temperature ($T$) dependence of resistivity ($\rho$) for as-grown and annealed FeS$_x$Se$_{1-x}$ films is displayed in Fig. 3(a). As reported previsouly,[20] thickness of 15 nm for FeSe is not thick enough to achieve metallic conduction and superconducting transition behavior in the as-grown state ($x$ = 0 in left panel); annealing induces metallic transport (right panel). A weak but definite $\rho$ decrease at low $T$ in the annealed FeSe film can be considered as the onset of superconductivity. The relatively low onset superconducting critical temperature ($T_c$) of about 5 K and incomplete superconducting transition are commonly observed for such thin films.[7,8,20] In FeS$_x$Se$_{1-x}$ alloy films ($x$ >0), $\rho$-$T$ curves are metallic with relatively low $\rho$ at high $T$ in the as-grown state, showing only a slight $\rho$ upturn at low $T$. However, we did not detect the onset behavior in all as-grown samples. In the annealed FeS$_{0.16}$Se$_{0.84}$ film, we observed an onset $T_c$ of ~ 3 K under zero magnetic field (blue line in the inset of right panel in Fig. 3(a)). The suppression of onset behavior as observed under 9 T is characteristic of superconductivity. The thermally activated transport in the as-grown FeS$_{0.89}$Se$_{0.11}$ and annealed FeS$_{0.44}$Se$_{0.56}$ are likely caused by disorder, consistent with the decomposition as revealed by XRD (Fig. 2).

Having observed the onset behavior in the annealed FeS$_x$Se$_{1-x}$ films ($x$ = 0 and 0.16), we examined the role of annealing by Hall effect measurement. Figure 3(b) displays $x$ dependence of Hall coefficient ($R_H$) at 50 K where normal state transport can be measured. In FeSe ($x$

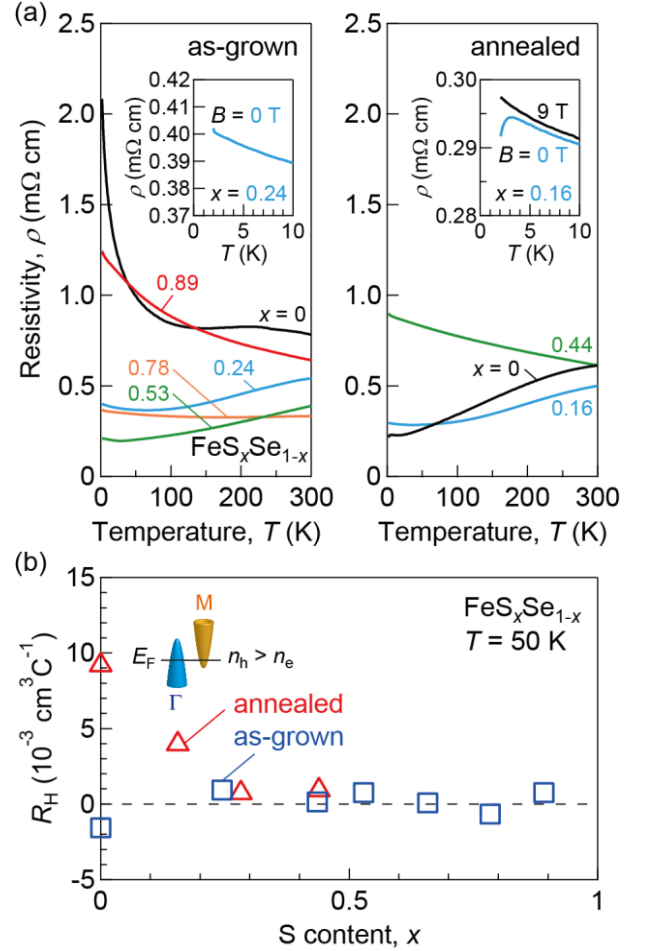

**Fig. 3.** (a) $\rho$-$T$ characteristics of as-grown (left) and annealed (right) FeS$_x$Se$_{1-x}$ films. Insets display the magnetic-field effect on the resistance decrease (superconductivity) at low temperatures. (b) $x$ dependence of $R_H$ at 50 K. A schematic band structure for annealed FeSe is presented in the inset.

= 0), the negative $R_H$ in the as-grown state becomes positive after annealing. According to the multiband model widely accepted for FeSe (hole pocket at Γ point and electron pocket at M point as depicted in Fig. 3(b) inset),[9] the $R_H$ variation corresponds to carrier-type crossover: effective hole carrier density ($n_h$) predominates electron carrier density ($n_e$), in association with the Fermi level shift. Previous works have shown that superconductivity in bulk FeSe and thick FeSe films favors slightly Fe-rich conditions[1,21,22] and positive $R_H$ values at low $T$. In fact, two annealed samples ($x$ = 0 and 0.16) exhibiting the onset $\rho$-$T$ behavior show positive $R_H$. On the other hand, the $R_H$ values for FeS$_x$Se$_{1-x}$ alloy films ($x$ > 0.2) are close to zero, and there are no apparent variations by annealing. This result implies that the hole and electron carriers balance in FeS$_x$Se$_{1-x}$ alloy films. Taking into account that metallic transport is already achieved in our alloy films with slightly Fe deficient compositions, it may be possible to induce superconductivity in the as-grown state by adjusting the stoichiometry, e.g., with Fe-rich FeS target.[1,21,22] Although there still remains such issues for further investigations, our films are expected to reflect different structural and electronic properties in high-$x$ FeS$_x$Se$_{1-x}$ alloy.

In summary, we have stabilized tetragonal FeS$_x$Se$_{1-x}$ up to $x = 0.78$ by alternate deposition using the PLD process. The effectiveness of alternate deposition in the expanded sulfur solid solubility would open a new way towards experiments on the FeSe – FeS alloy system.

**Acknowledgments**

We would like to thank I. Narita for his assistance with electron probe microanalysis. This work was performed under the Inter-University Cooperative Research Program of the Institute for Materials Research, Tohoku University (Proposal No. 17K0417). This work was partly supported by a Grant-in-Aid for Specially Promoted Research (No. 25000003) from Japan Society for the Promotion of Science.